\documentclass[twoside]{article}\textheight 9in\textwidth 6.5in
 \usepackage [T2A] {fontenc} \usepackage [utf8] {inputenc}
 \usepackage[russian,english]{babel} \usepackage{graphicx}
 \usepackage {amsmath,amssymb,amsthm} \usepackage [linktocpage] {hyperref}
\begin{document}

 \newtheorem* {prp} {Proposition} \newtheorem*{cor} {Corollary}

 \newif\ifpdf \ifx\pdfoutput\undefined \pdffalse
 \else\ifnum\pdfoutput=1 \pdftrue \else\pdffalse\fi\fi
 \newenvironment {pf}{Proof.}{}
 \newtheorem{thr} {Theorem} \newtheorem{dfn} {Definition}
 \newtheorem{rem} {Remark} \numberwithin{equation}{section}

 \author {Leonid A. Levin}
 \title  {Some Theorems on the Algorithmic Approach\\
          to Probability Theory and Information Theory.}
 \date   {\small Dissertation directed by A.N. Kolmogorov,
                 January 1971, translated by APAL in 2010.}

\maketitle\tableofcontents\vspace{1pc}

The dissertation uses the terms and notation in paper \cite{6} that is
attached to [Russian original of] this text. The paper also contains
figures referred to in the text of the dissertation as well as the
index.

The author is deeply grateful to his advisor {\sc A.\,N.\,Kolmogorov},
to {\sc A.\,K.\,Zvonkin} who helped a lot in presenting the results,
to {\sc V.\,N.\,Agafonov, Ya.\,M.\, Barzdin', R.\,L.\,Dobrushin,
A.\,G.\,Dragalin, M.\,I.\,Kanovich, A.\,N.\, Kolodiy,
P.\,Martin-L\"{o}f, L.\,B.\,Medvedovsky, N.\,Y.\,Pet\-ri,
A.\,B.\,Sosinsky, V.\,A.\,Uspensky, J.\,T.\,Schwartz}, and to all
participants of {\sc A.\,A.\,Markov's} seminar for discussion.

\subsection*{Some definitions and notation}
\addcontentsline{toc}{subsection}{Some definitions and notation}

We consider strings in the alphabet $\{0,1\}$, i.e., finite sequences of
zeroes and ones in a 1-1 correspondence with natural numbers:

\begin{eqnarray*} \Lambda &\leftrightarrow& 0\\ 0&\leftrightarrow&1\\
1&\leftrightarrow&2\\ 00&\leftrightarrow&3\\ 01&\leftrightarrow&4\\
10&\leftrightarrow&5\\ 11&\leftrightarrow&6\\ 000&\leftrightarrow&7\\
001&\leftrightarrow&8\\ \vdots&\vdots&\vdots\end{eqnarray*}
 ($\Lambda$ is the empty string). We do not distinguish strings and
number and use the terms interchangeably. They are usually denoted by
lower case Latin letters. The set of all strings-numbers is denoted by
$S$. The result of adding (concatenating) the string $y$ to the string
$x$ is denoted by $xy$. We need also to encode the ordered pair $(x,y)$
of strings by one string. To avoid introducing a special separator (such
as a comma) let us agree that for $x=x_{1}x_{2}\ldots x_{n}$
$(x_{i}\in\{0,1\})$
 \begin{equation}\overline{x}=x_{1}x_{1}x_{2}x_{2}\ldots x_{n}x_{n}01.
\label{01}\end{equation} Then one can recover both $x$ and $y$ from the
string $\overline{x}y$. Denote by $\pi_1(z)$ and $\pi_2(z)$ functions
such that $\pi_1(\overline{x}y)=x$, $\pi_2(\overline{x}y)=y$. If the
string $z$ is not representable as $\overline{x}y$ then
$\pi_1(z)=\Lambda$, $\pi_2(z)=\Lambda$.\footnote
 {More common enumerations of pairs may violate the property \eqref{011}
important below.}

The length $l(x)$ of a string $x$ is the number of its digits;
$l(\Lambda)=0$. Obviously
 \begin{equation} l(xy)=l(x)+l(y),\label{02}\end{equation}
 \begin{equation} l(\overline{x})=2l(\overline{x})+2.
\label{03}\end{equation}
 Let $d(A)$ be the number of elements in a set $A$. Obviously
 \begin{equation} d\{x:l(x)=n\}=2^n, \label{04}\end{equation}
 \begin{equation} d\{x:l(x)< n\}=2^n-1. \label{05}\end{equation}

We also consider the space $\Omega$ of infinite binary sequences,
denoting them with lower-case Greek letters. $\Omega^*=\Omega\bigcup
S$ is the set of all finite and infinite sequences. Let
$\omega\in\Omega^*$. The $n$-prefix of $\omega$, denoted $(\omega)_n$,
is the string of its first $n$ digits. If $\omega\in S$ with
$l(\omega)\le n$ then $(\omega)_n=\omega$ by definition. An
$\omega\in\Omega$ is a characteristic sequence for the set $S_\omega=
\{n_1,n_2,\dots\}$ of positive integers if $\omega$ has 1 at the
places $n_1,n_2,\ldots$ and zeroes everywhere else.
 Denote $\Gamma_x$ the set of all sequences (from $\Omega$ or $\Omega^*$,
as follows from the context) that have prefix $x$: \begin{equation}
\Gamma_x=\{\omega:(\omega)_{l(x)}=x\}. \label{06}\end{equation}

\begin{figure}\newcommand\ris{ris.eps}\ifpdf\renewcommand\ris{ris.pdf}\fi
\centering \caption{} \includegraphics{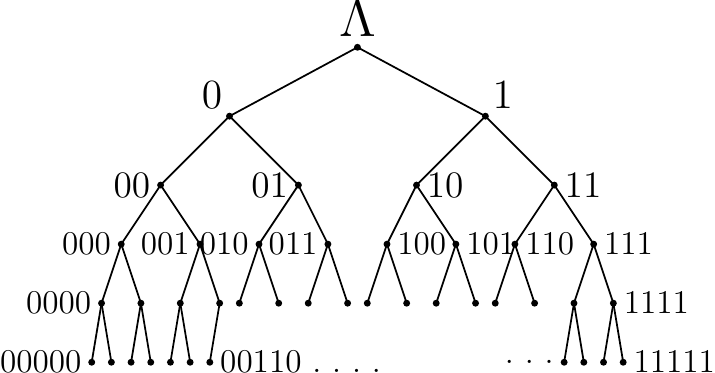} \end{figure}

Notation $x\subset y$ means $\Gamma_x \supseteq \Gamma_y$, that is the
string $x$ is a prefix of $y$. The relation $\subset$ is a partial order
on $S$ (Figure 1).

Functions defined on the Cartesian product $S^n=S\times \ldots \times S$
($n$ times) are denoted by capital Latin letters (except some standard
functions). Sometimes a superscript $n$ denoting the number of variables
is added: $F^n=F^n(x_1,\ldots,x_n)$. The sentence \emph {for all
admissible values of variables $y_1,\ldots,y_n$ there exists a constant
$C$ such that for all admissible values $x_1,\ldots,x_n$}
\begin{equation} F^{n+m}(x_1,\dots,x_n;y_,\dots,y_m)\le
G^{n+m}(x_1,\dots,x_n;y_,\dots,y_m)+C \label{07}\end{equation}
is abbreviated as follows: with parameters $(y_1,\ldots,y_n)$,
 \begin{equation} \label{08} F^{n+m}(x_1,\dots,x_n;y_,\dots,y_m)
\preccurlyeq G^{n+m}(x_1,\dots,x_n;y_,\dots,y_m).\end{equation}

The relation $\succcurlyeq$ is defined similarly. $F\asymp G$ means both
$F\preceq G$ and $G\preccurlyeq F$ hold. Obviously, the relations
$\preccurlyeq$, $\succcurlyeq$, $\asymp$ are transitive and
 \begin{equation} l(x)\asymp \log_2(x) \mbox{ for } x>0,
\label{09}\end{equation}
 \begin{equation} l(\overline{x})\asymp 2l(x),\label{010}\end{equation}
 \begin{equation} l(\overline{x}y)\asymp l(y) \qquad (\mbox{with $x$
as a parameter}).\label{011}\end{equation}

\section {Introduction}

\subsection {The general construction of complexity}

The topics studied here were introduced in 1964 when A.N.Kolmogorov
defined complexity of constructive objects. (Similar concepts were
independently considered by A.A.Markov and R.J.Solomonoff.)

A.N. Kolmogorov defines the complexity of a string $x$ for an algorithm
$A$ as the least length of binary strings $p$ encoding $x$, i.e., such
that $A(p)=x$. The value so defined depends strongly on the choice of
$A$. The central result that prompted all further investigations was a
theorem established by A.N. Kolmogorov and independently (in slightly
different terms) by R.J. Solomonoff. It states the existence of an
optimal algorithm $A$ providing the smallest (compared to any other
algorithm $B$) value of complexity up to an additive constant $C_B$
(independent of $x$). Complexity for an arbitrary optimal $A$ is thus
sufficiently invariant to be a fundamental characteristics of $x$. It
found many applications, and quickly generated a rich theory (cf. for
example, a survey [6]).

In the development of this theory, several other quantities similar to
complexity (though different from it) turned out to be useful. For
example, A.A.Markov and D.Loveland considered the decision complexity of
binary strings, P.Martin-L\"{o}f defined their deficiency of
``randomness,'' the present author introduced ``universal probability,''
etc. At present, about ten such functions are known. The need exists for
some organization of this diversity of quantities from a unified
standpoint.

\begin{dfn} A \emph {finitary function} is a table defining a
function from a finite set $A\subset S$ to $S$. (We assume
it has value $\infty$ on $A\backslash S$.) \end{dfn}

\begin{dfn} A \emph {volume restriction} is an enumerable
family $V$ of finitary functions such that\begin{enumerate}
 \item If $f\ge g$ and $g\in V$ then $f\in V$;
 \item $\exists C \forall f,g{\in}V$ $(C+\min \{f,g\})\in V$.
\end{enumerate} We assume for simplicity that $C=1$. \end{dfn}

\begin{dfn} Let $V$ be a volume restriction.
 By \emph {$V$-majorant}, we call any function $F(x)$ such that
 \begin{enumerate}
 \item the set of points over its graph is enumerable and
 \item for every finitary function $g$, if $g\ge F$ then $g\in V$.
\end{enumerate} \end{dfn}

\begin{thr} For any volume restriction $V$, there exists a $V$-majorant
$K_V(x)$, that is smallest (up to an additive constant), i.e., such
that $K_V(x)\preccurlyeq L(x)$ for every $V$-majorant $L(x)$. \end{thr}

\begin{pf} For a finite set $\cal M$ of pairs of numbers, we get a graph
of a finitary function by taking the lowest point of $\cal M$ on each
vertical line intersecting $\cal M$. Let us call this function a lower
boundary of $\cal M$.

Let partial recursive function $U(i,t)$ enumerate the $i$-th enumerable
set $U_i$ of pairs $(x,a)$ for every $i$. Let's define $U'(i,t)$
enumerating $U'_i\subset U_i$ for each $i$, but slower than $U$. Namely,
$U'$ generates the next element only after verifying that the lower
bound of the set enumerated so far belongs to $V$.

Obviously $U'_i$ is a $V$-majorant for each $i$, and no majorant is
``forgotten.'' Now let $\cal M$ be the set of pairs situated above pairs
$(x,a+Ci)$ where $C$ is the constant in the definition of volume
restriction, and $(x,a)\in U'_i$.

Let us prove that $\cal M$ defines an (obviously optimal) $V$-majorant.
In other words, every finitary function $f$ whose graph is contained in
$\cal M$, belongs to the family $V$. By definition of $\cal M$, $f\ge
\min(g_i+Ci)$ for some family of functions $g_i\in V$, $i\le n$.
 This implies that $f\in V$. Indeed let $ h_k=\min_{i>k}(g_i+C(i-k))$
 Then $f\ge h_0$, $h_{k-1}=C+\min \{h_k,g_k\}$ and induction on $k$
from $n$ down completes proving the theorem. \end{pf}

For any decidable volume restriction $V$, one can compute a common lower
bound $m_V(x)=\min_{f\in V} f(x)$ for all $V$-majorants. It is simpler
to study differences $K_V(x)-m_V(x)$ instead of the majorants $K_V(x)$.
These differences will be $V'$-majorants where $f\in V' \leftrightarrow
(f+m_V)\in V$. Obviously $m_{V'}(x)=0$. We call such $V'$ ``reduced.''
There is no need to study non-reduced decidable $V$.

\begin{thr} Among reduced volume restriction, there is the one that is
most ``narrow.'' The universal majorant $p(x)$ corresponding to it will
be the largest\footnote
 {This majorant is a logarithm of the largest (up to a constant)
semicomputable probability distribution on natural numbers}.
 This restriction is given by the condition $f\in V \Leftrightarrow
\sum_x 2^{-f(x)}\le 1$. \end{thr}

\begin{pf} Clearly, $V$ is a reduced volume restriction. If $V'$ is any
volume restriction and $f\in V$, then finitely many applications of item
2 of the definition of volume restrictions yield $f\in V'$. This
``extreme'' majorant $p(x)$ turns out to be not far from the complexity
$K(x)$ of [9] (which hence is close to the limit). \end{pf}

\begin{thr} $K(x)\preccurlyeq p(x)\preccurlyeq K(x)+2\log_2K(x)$.\end{thr}

\begin{pf} $K(x)\preccurlyeq p(x)$ by Theorem 2 (see also Theorem 4a).
 To prove the second inequality, we show that any finitary function
$f(x)\ge K(x)+2\log_2K(x)$ belongs to volume restriction $V$ (from
Theorem 2). Indeed,
 \[\sum_x2^{-f(x)}= \sum_a\sum_{x: K(x)=a} 2^{-f(x)} \le
\sum_a\sum_{{x: K(x)=a}} 2^{-K(x)-2\log_2K(x)}=\]
 \[=\sum_a d\{x:K(x)=a\}\cdot\frac{1}{2^a\cdot a^2}.\]
 Since $d\{x:K(x)=a\}\le2^a$, this $\le\sum_a \frac{2^a}{2^a\cdot
a^2}\preccurlyeq 1$, which completes the proof. \end{pf}

\subsection{Examples of majorants}

\begin{dfn} ({\sc A.\,N.\,Kolmogorov})\\
 The \emph {complexity} of $x$ with respect to a p.r.\ function $F^1$ is
\begin{equation*} K_{F^1}(x)\stackrel{\mathrm{def}}{=}\left\{
\begin{array} {l}\min_{F^1(p)=x}l(p)\\ \infty, \mbox{ if there is
no such }p.\end{array}\right. \end{equation*}
We call a word $p$ with $F^1(p)=x$ a code or program for $F^1$ to
restore $x$. \end{dfn}

\begin{dfn} ({\sc A.\,N.\,Kolmogorov})
 The \emph{conditional complexity} of $x$ for known $y$ with respect
to a p.r.\ function $F^2(p,y)$ is
 \begin{equation*} K_{F^2}(x\big|y)\stackrel{\mathrm{def}}{=}\left\{
\begin{array} {l}\min l(p):{F^2(p,y)=x}\\ \infty, \mbox{ if } \forall
p\,F^2(p,y)\neq x. \end{array}\right. \end{equation*} \end{dfn}

\begin{dfn} {(D.\,Loveland, A.\,A.\,Markov)}\\ The \emph{decision
complexity} of a word $x$ with respect to a p.r.\ function $F^2$ is
\begin{equation*} K_{F^2}(x)\stackrel{\mathrm{def}}{=}\left\{
\begin{array} {l}\min l(p):{\forall i<l(x)\,F^2(p,i)=x_i}\\ \infty,
\mbox{ if there is no such a }p, \end{array}\right.
\end{equation*} here $x_i$ is the $i$-th letter of word $x$.\end{dfn}

We defined three quantities important in complexity theory.
Let us show that all three are special cases of the general concept
of $V$-majorant. Then, in particular, Theorem 1 will imply the famous
optimality theorems discovery of which by A.\,N.\.Kolmogorov and
R.J.Solomonoff started complexity theory.

Let $V_1$ be the set of finitary functions with $\le2^a$ of $x$ having
$f(x)< a$.

Let $V_2$ be the set of finitary functions $f(x,y)$ on pairs $(x,y)$
(more precisely, on codes of such pairs) with every $a, y$ having
$\le2^a$ of $x$ with $f(x,y)< a$.

Let $V_3$ be the set of finitary functions with $\le2^a$
branches in the tree of words $x$ with $f(x)< a$.
 It is easy to verify that $V_1,V_2,V_3$ are volume restrictions.

We call classes $A$ and $B$ \emph {equivalent} if for every function $f$
from one of them there is a function $g\preccurlyeq f$ from the other.

\begin{thr} a) The class of $V_1$-majorants is equivalent to the complexity
class with respect to any algorithms.

b) The class of $V_2$-majorants is equivalent to the conditional complexity
class with respect to any algorithms.

c) The class of $V_3$-majorants is equivalent to the class of decision
complexities. \end{thr}

\begin{pf} We prove Theorem 4a. Theorems 4b and 4c can be proven similarly.
 It is easy to see that for any $A$, $K_A(x)$ is a $V_1$-majorant.
Conversely, for any $V_1$ majorant $F$ one can enumerate all points
$(x,n)$ above its graph and map them to different $p\in\{0,1\}^n$, with
pairs $(p,x)$ forming the graph of an algorithm $A$. $F\in V_1$ assures
enough codes $p$ for that. $K_A(x)$ may exceed $F(x)$ by $\le1$, QED. \end{pf}

\subsection{Invariant functions and complexity}

Complexity has an important property of invariance, namely

\begin{rem} Under any {p.r.} isomorphism between two recursively
enumerable sets, the complexities of their elements differ at most by a
constant.\end{rem}

Besides, complexity has ``informational correctness,'' namely

\begin{rem} There exists a computable enumeration of pairs $f(x,y){=}i$,
$\pi(i){=}x$, $\pi_2(i){=}y$ such that the complexity of $f(x,y)$ is at
least the complexity of $x$ and $y$ up to an additive constant (true
even for every computable enumeration). \end{rem}

Complexity is bounded by a logarithm of its argument, i.e.\ contains
only a very limited amount of information about the words. But it turns
out that even among functions of arbitrary nature, no ``richer''
invariants exist.

\begin{thr} Every invariant informationally correct (in the above
sense) function $F(x)$ is at most $K(x)$ up to a multiplicative
constant. The constant cannot be made additive because even changing the
alphabet changes a constant factor. \end{thr}

\begin{pf} The algorithm $A$ in $K_A(x)$ can be represented as the
composition of function $\pi_1(x)$ and an invertible function. Then
the theorem's assumptions imply $F(x)\preccurlyeq F(p)$ when $A(p)=x$.
 It remains to show $F(p)\le C\cdot l(p)$. This follows from constructing
four isomorphisms of natural numbers, combining which we can, in $n$ steps,
obtain every $n$-bit word from $0$. QED. \end{pf}

\subsection{Computable complexity majorants}

Clearly, knowing a word $x$ and its complexity, we can find efficiently
(at least, by exhaustive search) a shortest program coding $x$.
Moreover, knowing $x$ and any bound $S>K(x)$, we can find an $S$-bit
program, possibly not a shortest one. Since complexity isn't a
computable function, in practice one has to be content with its
computable majorants giving the length of an effectively computable
code, not necessary the shortest one. {\sc Barzdin'}, {\sc Petri}, {\sc
Kanovich} showed all such majorants to be very coarse in some cases.
However, we have

\begin{thr} Every ``informationally correct'' (in the sense of Sec. 1.3)
function which is less (up to an additive constant) than every
computable complexity majorant is also less (up to an additive constant)
than the complexity itself. \end{thr}

\begin{pf} Every algorithm can be represented as a composition of an
invertible algorithm assuming all values in a recursive set and function
$\pi_1(x)$. Complexity with respect to an invertible function is just the
logarithm of its inverse and hence it is computable. The theorem
follows. \end{pf}

\subsection{Decision complexity}

By many reasons, $K(x)$ or $K(x|l(x))$ are not quite natural to use
for studying the complexity of sequences (rather than terminated words).
Thus A.\,A.\,Markov and D.\,Loveland introduced $KR(x)$, which proved to
be very fruitful. E.g.,

\begin{rem} A sequence $\omega$ is computable if and only if
$KR((\omega)_n)$ is bounded. \end{rem}

Evident for $KR(x)$, this isn't true for $K(x)$, and for $K(x\big|
l(x))$ isn't evident and remained an open problem for some time. An
affirmative answer was given by the author independently of {\sc
Kolodiy}, {\sc Loveland} (USA) and {\sc Mishin}. This is implied by the
following theorem relating $KR(x)$ with $K(x|l(x))$.

\begin{thr} For every $\omega$, $KR((\omega)_n)$ is bounded if and only
if $K((\omega)_n|n)$ is.\footnote
 {However, it was shown by {\sc Petri} that there is no effective way to
calculate a bound on $KR((\omega)_n)$ from a bound on $K((\omega)_n|n)$,
that is, the former can be very large.} \end{thr}

\begin{pf} One direction is obvious: a computable $\omega$ has a general
recursive function $F^1(n)=(\omega)_n$. Let $F^2(p,n)=F^1(n)$, then
$K_{F_2}((\omega)_n\big|n)=l(\Lambda)=0$ because $F^2(\Lambda,n)=
(\omega)_n$, hence $K((\omega)_n\big|n)\preccurlyeq 0$,
$K((\omega)_n\big|n)\le C.$

Let us prove the other direction. Suppose $K((\omega)_n\big|n)\le C$. We
want to establish the existence of a procedure which, for given $n$,
produces $\omega_n$, the $n$-th digit of $\omega$. Consider all words
$p$ with length at most $C$ and construct a table as shown on Figure in
Sec. 2.2 of \cite{6}: at the $p$-th row of the $n$-th column we place
$F_0^2(p,n)$ (see (1.6) from \cite{6}) provided it halts. The set of all
words $F_0^2(p,n)$ in the $n$-th column we denote $A_n$. Each $A_n$ has
at most $2^{C+1}$ words, and $(\omega)_n\in A_n$.

Let $l=\overline{\lim_{n\rightarrow \infty}}\,d(A_n)$. Clearly,
the set $U=\{n:d(A_n)\ge l\}$ is recursively enumerable and infinite.
Moreover, there are only finitely many $n$ with $d(A_n)>l$;
the largest of such $n$ we denote $m_1$.
 Let $k<2^{C+1}$ be the number of sequences $\omega$ with
$K((\omega)_n\big|n)\le C$. Let $m_2$ be the smallest length of prefixes
distinct for all these sequences (by the way, all columns starting from
$m_2$-th should contain at least $k$ prefixes of these sequences, hence
$k\le l$). Let $m=\max(m_1,m_2)$.\footnote
 {Our construction uses numbers $l,k,m$ but isn't effective, giving no
procedure to find them. We only prove that the required algorithm exists
(an intuitionist would say: ``cannot but exist''), so only need the mere
existence of $l,k,m$.}

Let $U'$ be an infinite decidable subset of $U$ and $V=U'\cap\{n:n>m\}$.

The algorithm deciding the $i$-th (in lexicographical order) of our
sequences proceeds as follows. To find its $j$-th digit, we select the
smallest $n_r>j$ in $V$ and start filling in the $n_r$-th column (with
words $F_2(p,n_r)$, $l(p)\le C$). When $l$ words are found, we stop:
there are no more. Denote $B_{n_r}$ the set of all $n_r$-bit words from
$A_{n_r}$. Then we similarly construct the set $B_{n_{r+1}}$ and take
from it all words with prefixes from $B_{n_r}$; this set is denoted
$C_{n_{r+1}}$. Then, words from $B_{n_{r+2}}$ with prefixes from
$C_{n_{r+1}}$ form the set $C_{n_{r+2}}$; $C_{n_{r+3}}$ is be the set of
words from $B_{n_{r+3}}$ with prefixes from $C_{n_{r+2}}$, and so on.
 We stop when the current set $C_{n_s}$ contains exactly $k$ words: they
all are $n_s$-prefixes of sequences with $K((\omega)_n\big|n)\le C$.
Selecting the $i$-th lowest of them we take its $j$-th digit; it is what
is required. \end{pf}

\section{Measures and Processes}

This chapter considers deterministic and non-deterministic processes
generating sequences. The central result is introducing a universal
semi-computable measure and establishing its relation with complexity.
At the end of the chapter, these results are applied to the study of
capacities of probabilistic machines.

\subsection{Definitions. Equivalence of measures.}

\begin{dfn}\label{dfn2.1.1}\rm \emph{Algorithmic process} or simply
\emph{process} is a partial recursive function $F$ mapping words into
words, and such that if $F(x)$ is defined for a word $x$ and
$y\subset x$ then $F(y)$ is also defined and $F(y)\subset F(x)$.
\end{dfn}

Let us and apply a process $F$ to all prefixes of $\omega\in\Omega$ while
$F$ is defined. It outputs prefixes of a sequence $\rho\in\Omega^*$.
\footnote {If $F((\omega)_n)$ is defined, and for all $m>n$, $F((\omega)_m)$
coincides with $F((\omega)_n)$ or is undefined, then $F(\omega)=F((\omega)_n)$.
$F(\omega)=\Lambda$ if $F((\omega)_n)$ is undefined or empty for all $n$.}
 This $\rho$ is the result $F(\omega)$ of applying $F$ to $\omega$ (i.e., $F$
maps $\Omega$ into $\Omega^*$).

\begin{rem}\label{rmrk2.1.1.}\rm There exists a universal process,
i.e., a partial recursive function $H$, such that $H(i,x)$ is a process
for all $i$, and for any process $F$ an $i$ exists such that
$H(i,x)\equiv F(x)$. Such $H$ is easily constructed from a universal
p.r. function. Without loss of generality we assume (and use later)
$H(\Lambda,\Lambda)=\Lambda$.\end{rem}

Processes $F$ and $H$ are said to be \emph{equivalent} if
$F(\omega)=G(\omega)$ for any $\omega\in\Omega$.

\begin{rem}\label{rmrk2.1.2} Any process has an
equivalent one that is primitive recursive. \end{rem}

\begin{dfn}\label{dfn2.1.2}\rm We say a process $F$ is \emph
{applicable} to $\omega$ if $F(\omega)$ is infinite.\end{dfn}

\begin{rem}\label{rmrk2.1.3}\rm Any process is a continuous function
on the set of sequences to which it is applicable (with the natural
topology on $\Omega$).\footnote
 {In this topology, $\Omega$ is equivalent to Cantor perfect set.}
\end{rem}

\begin{dfn}\label{dfn2.1.3}\rm A process is \emph{fast growing} (\emph
{fast applicable to $\omega$}) if a monotone unbounded total recursive
function $\Phi(n)$ exists such that for all $x$ (respectively, for all
prefixes $x$ of $\omega$) for which $F$ is defined, $\ell(F(x))\ge\Phi(\ell
(x))$. In this case we say \emph {the speed of growth (of the
application to $\omega$) of process $F$ is $\ge\Phi(n)$}.\end{dfn}

\begin{rem}\label{rmrk2.1.4}\rm One can easily show that a process
applicable to all $\omega$ is total recursive and fast growing. Clearly,
the reverse is also true. \end{rem}

\begin{dfn}\label{dfn2.1.4}\rm Let $P$ be a probability measure over
$\Omega$. We say that process $P$ is \emph{regular} if the set of sequences
to which it is applicable has $P$-measure $1$. \end{dfn}

In order to define an arbitrary measure on a Borel $\sigma$-algebra of
subsets of $\Omega$, it suffices to define it on sets $\Gamma_x$.

\begin{dfn}\label{dfn2.1.5}\rm A measure $P$ on $\Omega$ is
\emph{computable} if there exist total recursive functions $F(x,n)$ and
$G(x,n)$ such that the rational number $\alpha_P(x,n)=\frac{F(x,n)}
{G(x,n)}$ is a $2^{-n}$-approximation of $P(\Gamma_x)$. \end{dfn}

\begin{rem}\label{rmrk2.1.5}\rm Obviously then, $\alpha_P(x,n+1)+2^{-n+1}$
is a $2^{-n}$-approximation of $P(\Gamma_x)$ from above. Hence, without
loss of generality, we always assume $\alpha_P(x,n)$ to be an upper bound,
and take $\alpha_P(x,n)-2^{-n}$ as a lower bound. \end{rem}

Denote by $L$ the measure $L(\Gamma_x)=2^{-l(x)}$, and call it the \emph
{uniform measure}. It corresponds to Bernoulli trials with probability
$1/2$; it is also a Lebesgue measure on the interval $[0,1]$. Obviously,
$L$ is computable.

\begin{thr}\footnote
 {A somehow weaker result was independently proven by Mann (USA).}
 a) For any computable measure $P$ and any $P$-regular process $F$ the
measure $Q(\Gamma_y)=P(\bigcup\,\Gamma_x:(F(x)\supset y))$ (i.e., the
measure with which the outputs of $F$ are distributed) is computable.

b) For any computable measure $Q$ there exists an $L$-regular process
$F$, generating $Q$-distributed outputs from $L$-distributed inputs.
Moreover, $F$ has an inverse $G$ (i.e.\ $F(G(\omega))=\omega$ when $G$ is
applicable) applicable to all non-recursive sequences except
maybe some in intervals of $Q$-measure $0$.\end{thr}

\begin{pf} a) We compute a $2^{-n}$-approximation (from above or below;
making it an upper bound is easy) $\alpha_Q(y,n)$ to $Q(\Gamma_x)$. Choose $m$
such that $P(\{\omega:l(F((\omega)_m ))>l(y)\})>1-2^{-(n+1)}$ (Such an $m$
exists as process $F$ is $P$-regular, moreover one can effectively find
such an $m$). Take all words $x{\in}\{0,1\}^m$ such that $y\subset
F(x)$, and compute $\alpha_Q(y,n)$ as the sum of
$2^{-(m+n+1)}$-approximations to measures $P(\Gamma_x)$ of all these
$x$. Then the error is $\alpha_Q(y,n)-Q(\Gamma_y)\le 2^{-(n+1)}+2^m\cdot
2^{-(m+n+1)}=2^{-n}$ (as there are $<2^m$ of $x$).

b) We consider sequences $\omega{\in}\Omega$ as reals in $[0,\,1]$ (with
binary expansions $\omega$; the cases of binary rationals, where such
expansions have ambiguity, will be specially noted).
 Figure~3 in \cite{6} shows a distribution function $g$ that corresponds
to measure $Q$. As is well known, the random variable $g^{-1}(\xi)$
 is $Q$-distributed with $\xi$ uniformly distributed over $[0,\,1]$.
 Our construction is based on this idea.

I. A process $F((\alpha)_n)$ generates $Q$-distributed $g^{-1}(\alpha)$
 from $L$-distributed inputs. It takes upper $2^{-2n}$-approximations
$\alpha_Q(y,2n)$ of $Q(\Gamma_y)$ for each $y{\in}\{0,1\}^n$ and outputs
the longest common prefix of those $z{\in}\{0,1\}^n$ for which
 \setcounter{equation}{0}\begin{equation}\label{eq1}
 \sum_{y\le z}\alpha_Q(y,2n)\ge(\alpha)_n\ge
 1{-}2^{-n}-\sum_{y\ge z}\alpha_Q(y,2n).\end{equation}

II. Due to \eqref{eq1}, the intervals $\cup\Gamma_z$ contain
 (for each $n$) $g$-image of $\alpha$. Hence, $F(\alpha)$, if applicable,
generates $g^{-1}(\alpha)$ (treating $\gamma$ in Figure~3 of \cite{6}
as a pre-image of $\alpha\in[\sigma',\, \sigma'']$).
 To prove $F$ is $L$-regular suffices to show it being what we need.

1) Let $[\sigma',\,\sigma'']$ correspond to a single $\gamma$ with
$Q(\gamma){>}0$. If $\sigma'<\alpha<\sigma''$ then once $\sigma'\le
(\alpha)_n{-}2^{-n}\le(\alpha)_n{+}2^{1-n}\le\sigma''$, only a single $z$
satisfies \eqref{eq1} and so is output. Thus, $F$ is applicable to
 such $\alpha$, though not always to the ends $\sigma'$, $\sigma''$.

2) Now let $\alpha$ not be of such types. Then $Q(\cup\Gamma_z)\rightarrow
0$ as $n\rightarrow\infty$. Hence, if $\alpha$ is not of type $\rho$
corresponding to a measure $0$ interval, then $\cup\Gamma_z$ shrink to a
point $\beta=g^{-1}(\alpha)$; their longest common prefix grows infinitely.

3) A notable case of type $\rho$ is $\alpha=g(\beta)$
 with a binary rational $\beta$: its two binary expansions
 may form a measure $0$ interval mentioned above.

In sum, $F$ is applicable to all sequences except some of types
 $\rho$, $\sigma'$, $\sigma''$ of Figure~3 in \cite{6}.
 This set is clearly countable, so $F$ is $L$-regular.

III. The inverse process $G$ just computes $g$. It may be non-applicable
only to (computable by Corollary to Theorem~11) $\gamma$ with $Q(\gamma)
{>}0$, and $\beta$, with binary rational $\alpha=g(\beta)$. If $F(\alpha)$ is
applicable, it computes $\beta$. If not, and $\beta$ is not of mentioned
type $\gamma$, it lies on an interval $[\tau',\,\tau'']$ of zero
$Q$-measure. Q.E.D.\end{pf}

\subsection{Semi-computable measures}

\begin{dfn} \emph {A semi-computable} (the term is justified by
Theorem~9) measure is the distribution of the outputs of an arbitrary
(not necessarily regular) process on inputs distributed according to a
computable measure. \end{dfn}

\begin{rem} Semi-computable measures are concentrated on $\Omega^*$
since a non-regular process can have finite outputs with positive
probability. In this section, we assume $\Gamma_x$ is a set of all finite
and infinite sequences with prefix $x$.\end{rem}

\begin{rem} The distribution of outputs of any process on inputs with
an arbitrary semi-computable distribution is also semi-computable
(as a composition of two processes is again a process).
Any semi-computable measure can be obtained from a uniform
measure by some process (see Theorem~8b). \end{rem}

\begin{thr} A measure $P$ is semi-computable iff total recursive
functions $F,G$ exist such that $\beta_P(x,t)=\frac{F(x,t)}{G(x,t)}$
 is a monotone non-decreasing in $t$ function, and
 \begin{equation}\label{eq3} \lim_{t\rightarrow\infty}\,
\beta_P(x,t)=P(\Gamma_x).\end{equation}\end{thr}

This Theorem implies that the class of semi-computable measures
(more accurately, of their logarithms) is equivalent to the class of
$V$-majorant, where $V$ is a set of finitary functions $f$ for which
$\sum_{x{\in}M} 2^{-f(x)}\le1$ for all sets $M$ whose
words are not prefixes of each other.

{\bf Proof.} Let $P$ be a semi-computable measure. Then there exists a
process $F$ generating this measure from $L$. Let it make
$t$ steps on all words $y$ with $\ell(y)\le t$ and, denoting the result
by $F_t(y)$ (if no results are achieved yet then $F_t(y)=\Lambda$),
set $\beta_P(x,t)=L(\cup\,\Gamma_y:x\subset F_t(y))$.

Inversely, suppose a measure $P$ has a function $\beta_P(x,t)$
satisfying the terms of the Theorem. We wish to construct a process $F$
generating $P$ from $L$.

The idea is simple: we need to partition the interval $[0,\,1]$ into
disjoint subsets of measure $P(\Gamma_x)$, and to output $x$ when our
uniformly distributed input falls into a corresponding set. Now we
describe the construction precisely. Clearly, $P(\Gamma_x)\ge
P(\Gamma_{x0})+P(\Gamma_{x1})$. Moreover, without loss of generality, we
assume $\beta_P(x,t)\ge\beta_P(x0,t)+\beta_P(x1,t)$ for all $t$: each
time this fails, we delay growth of $\beta_P(x0,t)$ and $\beta_P(x1,t)$
proportionally to restore the inequality. It is easy to construct
subsets of interval $[0,1]$ with the following conditions: to each pair
$(x,t)$ there corresponds a union $I_{x,t}$ of a finite number of
intervals with binary rational ends and combined length $\beta_P(x,t)$.
Within this procedure for any words $x\ne y$ of equal length,
$I_{x,t_1}$ and $I_{y,t_2}$ are disjoint for all $t_1$ and $t_2$; for
any words $x\subset y$ and any $t$, $I_{y,t}\subset I_{x,t}$; for any
$t_1<t_2$ and any $x$, $I_{x,t_1}\subset I_{x,t_2}$.

Our $F(z)$ constructs $I_{x,t}$ for all $x,t$ such that $l(x)\le l(z)$
and $t\le l(z)$, and outputs a longest $x$ such that $z\in I_{x,t}$ for
some $t$. Obviously, such $x$ is unique as the sets corresponding to
divergent $x$ are disjoint, and $x'\subset x''$ for $z'\subset z''$.

\subsection{Universal semi-computable measure}

\begin{thr} There exists a semi-computable measure $R$ that is \emph
{universal,} i.e., such that for any semi-computable measure $Q$, there
is a constant $C$ such that $C\cdot R(\Gamma_x)\ge Q(\Gamma_x)$ for all
$x$.\footnote
 {In other words, $Q$ is absolutely continuous with respect to $R$ with
Radon-Nikodym derivative bounded by $C$ from above.} \end{thr}

\begin{pf} By a remark in Section 2.1, a universal process $H(i,x)$
exists. Obviously, $F(z)\stackrel{\rm def}{=}H(\pi_1(z),\pi_2(z))$ is a
process. Applied to uniformly distributed sequences, it generates the
desired measure. Indeed, let a process $G(x)$ ($=H(i,x)$ for some $i$
and all $x$) transform some set of sequences into $\Gamma_x$. Then
$F(x)$ transforms into $\Gamma_x$ these same sequences with added prefix
$\overline i$ -- (maybe some others as well). Thus, the measure of
$\Gamma_x$ cannot decrease by a factor $>C2^{\ell\left(\overline
i\right)}$ ($\approx i^2$). \end{pf}

\begin{rem} This result does not extend to computable measures:
no measure is universal among all computable measures. This is one of
the reasons for introducing the notion of a semi-computable measure.

The measure $R$, being (within a constant factor) ``larger'' than any
other measure, is concentrated on the widest subset of $\Omega^*$.
\end{rem}

The following issue is considered in mathematical statistics: find out
what distribution $P$ can randomly generate a given sequence $\omega$.
If we know nothing a priori about $\omega$, then the only (= the
weakest) statement we can make about it is that it can be generated
under distribution $R$. In this sense, $R$ reflects our intuition about
``prior probability.'' The following is of interest:
 \begin{itemize} \item[a)] For a constant $C$, the probability (under
measure $R$) of having a $1$ after $n$ zeros is
$>\frac{1}{n}\cdot\frac{1}{C\log^2n}$.
 \item[b)] For every constant $C$, at most $\frac{1}{C}$ fraction of $n$
on any interval $[0,N]$, has the probability (under $R$) of a $1$
falling after $n$ zeros to exceed $\frac{1}{n}\cdot C\log^2n$.
\end{itemize}
 Thus, $R(0^n1)$ is typically around $\tfrac{1}{n}$.\footnote
 {Note that this statement is true only for the universal (prior)
probability. For example, if we know that the Sun has risen for 10,000
years, this does not mean that the probability of the Sun not rising
tomorrow is approximately equal to 1/3,650,000. This statement would be
true if the above fact was the only information that we have about the
Sun.}

The proof easily follows from Theorem \ref{thm11}, taking into account
that the complexity $KR(0^n1)$ does not exceed $\log_2n+c$, and for the
majority of these words this complexity is almost equal to $\log_2 n$.

One can see an analogy between constructing the complexity $KR$ and the
universal semi-computable measure. It turns out these two quantities
also have a quantitative connection:

\begin{thr} \label{thm11} $|KR(x)-(-\log_2 R(\Gamma_x))|\preccurlyeq
2\log_2 KR(x).$ \end{thr}

\begin{pf} Let $KR(x)=i$, thus, for some $p{\in}\{0,1\}^i$ and all
$n\le\ell(x)$, we have $G^2_0(p,n)=x_n$ (here $G^2_0$ is from Theorem
2.1 of \cite{6}). Then, one can easily construct a process transforming
each sequence with prefix $\overline{\ell(p)}\,p$ into a sequence with
prefix $x$: this process first separates the prefix $\overline
{\ell(p)}$, recovers $\ell(p)$, ``reads'' $p$, and sequentially
generates $G^2_0(p,n)$ for $n=1,2,\ldots$. From a uniformly distributed
input, this generates sequences in $\Gamma_x$ with probability $\ge
2^{-\ell\left(\overline{\ell(p)}\, p\right)}$. Thus, by Theorem 10,
$R(\Gamma_x)\ge c\cdot 2^{-\ell(\overline{\ell(p)}\,p)}$, hence
 \[-\log_2 R(\Gamma_x)\preccurlyeq\ell\left(\overline{\ell(p)}\,
 p\right)=\ell(p)+2\ell(\ell(p))=i+2\ell(i)=KR(x)+2\ell(KR(x)).\]

Now, assume $R(\Gamma_x)=q$. Let us denote $\ell(q)=\lfloor -\log_2
q\rfloor.$ To estimate the complexity $KR(x)$, we reconstruct every
symbol of $x$ from the triple $\ell(q),k,i$ (i.e., from $\overline{\ell
(q)}\,\overline{k}\,i$), where $k{\in}\{0,1\}$ and $i\le 2^{\ell(q)+1}$.
Our algorithm works as follows: based on $\ell(q)$, it builds the tree
(see Fig.~4 in \cite{6}) of all words $y$ with $R(\Gamma_y)>
2^{-\ell(q)-1}$. For this, we compute $\beta_R(y,t)$ for more and more
values $t$ and $y$, and add $y$ to the tree when we get
$\beta_R(y,t)>2^{-\ell(q)-1}$, for some $t$.

The word $x$ belongs to this tree. We keep only ``maximal'' words, i.e.,
words that are not prefixes of other words in the current tree. Clearly,
the number of such ``maximal'' words will neither decrease nor exceed
$2^{\ell(q)+1}$. Let $A$ (see Fig.~4 from \cite{6}) be the word from
which the last branching from the word $x$ occurs; after this, $x$
continues without branching. To find $x$, it suffices to have the first
digit $k$ of $x$ extending $A$, and the number $i$ of maximal words at
the moment when the tree being constructed branches at $A$ (incrementing
the number of maximal words to $i$). As $i\le 2^{\ell(q)+1}$, hence
$\ell(i)\le \ell(q)+1$. Thus,
 \[KR(x)\preccurlyeq\ell\left(\overline{\ell(q)}\,\overline{k}\,i\right)
 \asymp 2\ell(\ell(q))+\ell(i)\preccurlyeq2\ell(\ell(q))+\ell(q)\asymp\]
 \[-\log_2 R(\Gamma_x)+ 2\log_2(-\log_2 R(\Gamma_x)).\]
 But, as proven earlier, $2\log_2(-\log_2 R(\Gamma_x))\preccurlyeq
 2\log_2[KR(x)+2\ell(KR(x))]\preccurlyeq 2\log_2 KR(x)$,
so $KR(x)\preccurlyeq -\log_2 R(\Gamma_x)+2\log_2 KR(x)$.
 The theorem is proved. \end{pf}

\begin{rem} It is worth mentioning that the usual measure-theoretic
arguments assure that each measure $P$ (not necessarily semi-computable)
is almost fully concentrated on the set of all sequences $\omega$ for
which $\exists c\,\forall n$ $P((\omega)_n)\ge c\cdot R((\omega)_n)$.

Similarly, for $R$-almost all sequences, the inverse inequality also
holds. If $P$ is absolutely continuous with respect to $R$, then the
inverse inequality also holds for $P$-almost all sequences. This implies
a statement similar to Theorem 11 for an arbitrary semi-computable
measure $P$ and prefixes of $P$-almost any sequence (of course, the
constants may vary with sequences). \end{rem}

As a corollary, we get the well-known de Leeuw-Moore-Shapiro-Shannon
theorem about probabilistic machines:

\begin{cor} A sequence is computable if and only if some semi-computable
measure (hence, also the universal measure) is positive on it.\end{cor}

\begin{pf} By Theorem \ref{thm11}, the measure of all
prefixes is larger than a positive number if and only if their
``complexity of solution'' $KR$ is uniformly bounded. \end{pf}

\subsection{Probabilistic machines}

The above Shannon et al. result is sometimes interpreted as the
impossibility for probabilistic machines to solve problems that are
unsolvable deterministically. However, not all problems require
constructing a specific unique object; some allow many solutions, being
satisfied by producing any of them\footnote
 {The corresponding concept of a mass problem was formulated in
\cite{31}.}.
 This class clearly has problems that are unsolvable by deterministic
machines but solvable if a machine can use a random number generator. An
example of such problems is: to generate an uncomputable sequence.

We say a problem of constructing a sequence with a property $\Pi$ is
\emph {solvable on a probabilistic machine} if the universal measure $R$
of the set of all such sequences is positive. The following theorem
shows that such problems can indeed be solved on machines with an access
to random number generators; moreover, they can be solved with an
arbitrary given reliability, and quite efficiently, i.e., with small
number of calls to the random number generator. We call functions $f(n)$
and $g(n)$ \emph {asymptotically equal,} denoted $f(n)\sim g(n)$, if
$\frac{f(n)} {\log_2 f(n)}\asymp \frac{g(n)}{\log_2 g(n)}$; in this
section, inequalities $\preccurlyeq$, $\succcurlyeq$ are understood
in a similar ``asymptotic'' way.

\begin{thr} \label{thm12} Let $A\subseteq\Omega$ with $R(A)>0$. Then,
for every $\varepsilon>0$, there exists a fast-growing process $F$
(i.e., one with $\ell(F(x))\succcurlyeq \ell(x)$) transforming
$L$-distributed sequences into sequences in $A$ with probability
$>1-\varepsilon$ .\footnote
 {Note that first, the construction of this process is not always
efficient and second, as shown by N. Petri, this process sometimes
cannot be replaced by a table-based one (i.e., by a total recursive
fast-growing process).} \end{thr}

Clearly, one cannot solve, e.g., the problem of obtaining a very complex
sequence by using a process which grows faster, since the process cannot
increase the complexity of words. If sequences from the desired set $A$
have small complexity, then short programs can generate prefixes of
these sequences. However, one can imagine that some $A$ could make such
programs so special that short random inputs cannot be used instead,
forcing the slow growth of processes solving $A$. Interestingly,
``fast'' processes are also possible in all such cases.

\begin{thr}\label{thm13} Let $g$ be a monotonic total recursive
function. Then the problem of generating a sequence from a set $A$ is
solvable by a random process that grows as $g(n')$ or faster ($n'\sim
n$), if and only if there exists a set\footnote
 {This set $B$ can always be selected to be closed.}
 $B\subseteq A$ such that $R(B)>0$ and all $\omega\in B$, have
$KR((\omega)_{g(n)})\preccurlyeq n$. \end{thr}

\begin{pf} The previous paragraph makes one direction obvious;
we prove the other. Let $B{\subseteq}A$, $R(B){>}0$ and
$KR((\omega)_{g(n)})< n+c\log n$ for all $\omega{\in}B,n$.

We first construct a semi-computable measure $P$ with $P(B){>}0$ and
{\bf integer} $P((\omega)_{g(n)})2^{t_n}$, for $t_n=n+\lceil O(1){+}
(c{+}4)\log n \rceil$. For that, we round down $R((\omega)_{g(n)})$ (cut
proportionally to satisfy $P(x0)+P(x1)\le P(x)$). Each rounding cuts the
measure by $<2^{-t_n}$. But $KR(x)< n+c\log n$ for $g(n)$-prefixes of
$\omega{\in}B$, hence by Theorem~\ref{thm11}, $R(x)=2^{-n}/O(n^{c+2})$.
Thus, each rounding cuts $O(1)/n^2$ fraction of their measure, leaving
$R((\omega)_{g(n)})/O(1)$ for $\omega{\in}B$, and $P(B)>0$.

We then construct a process generating $P$ as in the proof of Theorem~9,
but select sets corresponding to pairs $(x,t$) (where $\ell(x)=g(n)$) to
consist of intervals of length multiple of $2^{-t_n}$. Clearly this
process is the desired one. \end{pf}

Let us describe another result about solvability of standard algorithmic
problems on probabilistic machines. The first interesting result of this
type was proven by Janis Barzdin'. An infinite set of natural numbers is
called \emph {immune} if it does not contain any infinite recursively
enumerable subset.

\

\begin{prp} {(Barzdin') \em There exists an immune set for which the
problem of constructing a characteristic sequence of its infinite subset
is solvable by a probabilistic machine.} \end{prp}

\

The class of all immune sets contains an interesting subclass of
hyper-immune sets. For these sets, the following result holds.

\begin{thr}\label{thm14}\footnote
 {Also proven by V.N. Agafonov independently of the author.}
 For any hyper-immune set $M$, the problem of constructing a characteristic
sequence of its infinite subset cannot be solved by a probabilistic
machine. \end{thr}

\begin{pf} Assume a machine can solve this problem with a positive
probability. Then, by Theorem 12, a machine can solve it with
probability $p>2/3$. Construct a function $f(i)$ computed by the
following algorithm: run this machine on the tree of sequences until it
generates, on measure $\ge\frac{2}{3}$, some sequences that have at
least $i$ ones; return the maximum of the positions of these ones.
Clearly, this function dominates the direct enumeration of $M$. Q.E.D.
\end{pf}

Petri subsequently showed that if $M$ is not fixed, the problem of
generating the sequence characteristic for a hyper-immune set is
solvable on probabilistic machines. However, the following holds:

\begin{thr}\label{thm15} Let us call a set \emph {strongly hyper-immune}
if its direct enumeration dominates, from some place on, each computable
function. The problem of generating sequences characteristic for
strongly hyper-immune sets is not solvable on probabilistic machines.
\end{thr}

The proof is similar to the one above.

\section{Information Theory}

\subsection{Definition and basic properties}

The complexity $K(x)$ denotes, intuitively, the amount of information
required for restoring a text $x$. The conditional complexity $K(x|y)$
--- the amount of information needed in addition to the information in
text $y$, for restoring text $x$. The difference between these two can
be naturally called the amount of information in~$y$ about~$x$.

\begin{dfn} ({\sc A.N.Kolmogorov})

The \emph {amount of information in $y$ about $x$} is
$I(y:x)\stackrel{\rm def}{=}K(x)-K(x|y)$. \end{dfn}
 \begin{rem} (a) $I(x:y)\succcurlyeq 0$; (b) $I(x:x)\asymp K(x)$.\end{rem}

\begin{pf} (a) Let $F^2(p,x)=F^1_0(p).$ Then, if $F^1_0(p_0)=y$ and
$K(y)=\ell(p_0)$, from $F^2(p_0,x)=y$, we conclude:
$K(y|x)\preccurlyeq K_{F_2}(y|x)=K(y).$

(b) Let $F^2(p,x)=x$. Then $F^2(\Lambda,x)=x$, and
consequently $K(x|x)\preccurlyeq K_{F^2}(x|x)=\ell(\Lambda)=0.$ It
remains to observe that $I(x:x)=K(x)-K(x|x)$. \end{pf}

\subsection{Commutativity of information}

{\sc Shannon}'s classical definition of the amount of mutual information
in two random variables is commutative, that is, $J(\xi:\eta)= J(\eta:
\xi)$. For Kolmogorov's concept of the amount of information in one text
about another, such a precise equality, in general, will not hold.

\vspace{6pt}\noindent Example. Clearly, some words $x$ of each length
have $K(x|\ell(x))\ge\ell(x){-}1$.\\
 By Theorem 4(b), there exist arbitrarily large values of $l$ with
 $K(l)\ge \ell(l)-1.$ For so chosen pairs $x$, $l{=}\ell(x)$,
 \[\begin{array} l I(x:l)=K(l)-K(l|x)\succcurlyeq \ell(l),\\
I(l:x)=K(x)-K(x|l)\preccurlyeq l-l=0. \end{array}\]

Thus, in some cases, $I(x:y)$ and $I(y:x)$ differ by order of the
logarithm of the complexities of $x,y$. But {\sc A.N.~Kolmogorov} and
{\sc L.~Levin} showed independently in 1967 that this is the largest
possible order of magnitude for this difference. So, disregarding the
smaller order quantities, $I(x,y)$ is commutative. Specifically, {\sc
A.N.~Kolmogorov} and {\sc L.~Levin} proved the following:

\begin{thr} \label{thm16}\footnote
 {With more careful estimates, the bound can be tightened.
 For instance, $12$ can be replaced by $5{+}\epsilon$.
 It is not known whether it can be brought down to $1$.}
 \begin{itemize}
 \item[(a)] $|I(x:y)-I(y:x)|\le 12\ell(K(\overline{x}y))$;
 \item[(b)] $|I(x:y)-[K(x)+K(y)-K(\overline{x}y)]|
\le12\ell(K(\overline{x}y))$. \end{itemize} \end{thr}

\begin{pf} (a)~We prove the inequality in one direction:
\begin{equation}\label{e1} I(x:y)\succcurlyeq I(y:x)-12\ell(K(xy)).
\end{equation} The other follows by swapping $x$ and $y$.

We construct two auxiliary functions. Let the partial recursive function
$F^4(n,b,c,x)$ enumerate without repetitions the words $y$ such that
$K(y)\le b$, $K(x|y)\le c$. The existence of such a function follows
from \cite[Theorem 0.4]{6} (taking into account the remark). Let $j$ (an
uncomputable function of $x,b,c$) be the number of such words~$y$.
Function $F^4$ halts for all $n\le j$ and only for them. Hence the
predicate $\Pi(b,c,d,x)$, asserting that $j$ defined above is $\ge2^d$,
is equivalent to halting of $F^4(2^d,b,c,x)$ and so is partial
recursive. Similarly, there exists a function $G^5(m,b,c,d)$
enumerating without repetitions all words~$x$ with $\Pi(b,c,d,x).$
 Denote by $i$ (an uncomputable function of $b,c,d$) the number of such
$x$. Obviously $G^5(m,b,c,d)$ halts for all~$m\le i$ and only for them.

We now start the proof. Let $a=K(x)$, $b=K(y)$, $c=K(x|y)$.
 Then \[I(y:x)=a-c.\]
 With $j,d{=}\ell(j)$ and $i$ so defined, clearly $i\cdot 2^d$ does not
exceed the number of pairs $(x',y')$ with $K(y')\le b$, $K(x'|y')\le
c$. That number is $\le 2^{b+c+2}$, so
 \begin{equation}\label{e2} \ell(i)+\ell(j)\preccurlyeq b+c.\end{equation}
 Since $F^4(n,b,c,x)$ returns $y$ for some $n\le j$,
 \begin{equation} \label{e3} K(y|x)\preccurlyeq\ell(\overline b\overline
cn)\preccurlyeq 2 \ell(b)+2 \ell(c)+ \ell(j). \end{equation}
 Furthermore, since $G^5(m,b,c,d)$ returns $x$ for $d=\ell(j)$ and
some $m\le i$,
 \begin{equation}\label{e4}a=K(x)\preccurlyeq\ell(\overline b\overline c
\overline d m)\preccurlyeq2\ell(b)+2\ell(c)+2\ell(d)+\ell(i).\end{equation}

Inequalities (\ref{e2})--(\ref{e4}) and the $\ell(K(\overline xy))$
bound on each of $\ell(b),\ell(c),\ell(d)$ imply $K(y|x)\preccurlyeq
b+c-a+12\ell(K(\overline xy))$. Claim~(a) follows.

(b) Clearly, $K(\overline x\,\overline x y)\preccurlyeq K(\overline
xy)$. So, by claim~(a), $|I(\overline xy:x)-I(x:\overline xy)|\preccurlyeq
12\ell(K(\overline xy))$, that is, $|K(\overline xy)-K(\overline
xy|x)-K(x)+K(x|\overline x y)| \preccurlyeq 12\ell(K(\overline xy))$, or
 \[|[K(\overline xy)-K(x)-K(y)]+K(y)-K(\overline xy|x)+K(x|\overline xy)|
\preccurlyeq 12\ell(K(\overline xy)).\]
 Now claim (b) follows by noting that $K(x|\overline xy)\asymp 0$,
$K(\overline xy|x)\asymp K(y|x)$. \end{pf}

\subsection[Entropy of arbitrary dynamic systems]
{Entropy of arbitrary dynamic systems (stationary stochastic
processes) and algorithmic amount of information}

{\sc A.N.Kolmogorov} showed that for processes of independent trials the
algorithmic amount of information is asymptotically equal to the
classical (probabilistic) one (see \cite[Theorem 5.3]{6}). In view of
Theorem~\ref{thm16}(b), this follows from the connection between
algorithmic complexity and probabilistic entropy.

{\sc J.T.Schwartz} posed the question of whether a similar fact holds
for an arbitrary ergodic stationary process (that is, a process for
which entropy is defined). We give a positive answer to this question in
the following theorem.

\begin{thr} \label{thm17} Let $\{\xi_i\}$, $i=1,2,\dots$, be an
arbitrary ergodic stationary stochastic process with values
$\xi_i\in\Omega$, let~$P$ be the measure on its trajectories
$u\in\Omega^{\cal N}$ that defines this process, and let~$H$ be its
entropy. Denote by $\alpha^i_n(\omega)$ the word $(\overline{\xi_1})_n
(\overline{\xi_2})_n\cdots(\overline{\xi_i})_n$. Then for $P$-almost all
$u$ \[\lim_{n\rightarrow\infty} \lim_{i\rightarrow\infty}\ \frac{K(
\alpha^i_n(\omega))}i=H.\] \end{thr}

Clearly, the ergodicity requirement here is not essential. For
non-ergodic processes, instead of their average entropy~$H$, one would
take ``entropy at a point'': a function measurable with respect to
the $\sigma$-algebra of invariant sets, averaging on any such set to its
average entropy on the set. This easily follows ``decomposition'' of
arbitrary stationary stochastic processes into ergodic ones.

Returning to the ergodic case, it suffices to prove the theorem for
processes with discrete values $(\xi)_n$. The general case will follow
by taking the limit on~$n$.

Consider the set of $2^n$-ary sequences $u$ -- trajectories of our
stochastic process. Defined on this set is a transformation~$T$ shifting
the time by 1 and a $T$-invariant ergodic measure describing the
process. Within $k$ time steps, $2^{n\cdot k}$ different
sequences~$X^k_i$ of length~$k$ can appear. Clearly, for
every~$\epsilon$, a~$k$ exists such that
 \[-\sum_{i<2^{n\cdot k}}P(X^k_i)\log_2 P(X_i^k)\le k\cdot(H+\epsilon).\]
 Since $T^k$, as well as~$T$, preserves the measure~$P$, it follows from
the Central Ergodic Theorem (C.E.T.) that for $P$-almost every sequence
there exists, for every~$l$, a limit of the frequency of the values
of~$m$ for which the sequence $T^{mk+l}(\omega)$ begins with $X^k_i$.

Take any such~$\omega$ and denote these limits for it by $P_{i,l}$. From
C.E.T. for~$T$ and the ergodicity of~$T$, it follows that almost always
$\sum_{l\le k}\frac{P_{i,l}}k=P(X^k_i)$. Hence we have here~$k$
probability distributions on the finite set $X^k_i$ and their average
with an entropy $\le k(H+\epsilon) $.

By convexity of entropy, at least one of the summand distributions has
entropy $\le K(H+\epsilon)$. So for our~$u$, an $l$ exists such that the
entropy of the frequencies $P_{i,l}$ with which number~$m$ satisfies the
condition ``$T^{mk+l}(\omega)$ begins with $X^k_i$'' is $\le
K(H+\epsilon)$. By a theorem of Kolmogorov (see \cite[Theorem 5.1]{6}),
it follows that the ``unit complexity'' of almost all~$u$ is $\le H$,
which gives a ``half'' of our theorem.

To prove the unit complexity to be $\ge H$, we use some results from
Section~2. Consider the collection $X^k_i$ of values of some
realization~$\omega$ of the process over the first~$k$ time steps and
compare four quantities: the entropy~$H$; the logarithm of the
probability of that collection divided by $k$, that is, $\frac{\log
P(X^k_i)}k$; the logarithm of its a priori probability (see the
definition of~$R$) divided by~$k$ also, that is, $\frac{\log
R(X^k_i)}k$; and the unit complexity $\frac{K(X^k_i)}k$. Their limits as
$k\rightarrow\infty$ are equal. For the first two quantities, this
follows from the Shannon-McMillan-Breiman theorem; for the last two,
from Theorem 11 of this dissertation; and for the two in the middle,
from the last remark of Section 2.3. The theorem is proved.

\end{document}